\documentclass[11pt,superscriptaddress]{revtex4}

\usepackage{picinpar}
\usepackage{floatflt}
\usepackage{array}
\usepackage{subfigure}
\usepackage{amsmath}
\usepackage[dvips]{graphicx}
\usepackage{amssymb}

\DeclareSymbolFont{operators}{OT1}{cmss}{m}{n}
\DeclareSymbolFont{letters}{OML}{cmss}{m}{it}
\DeclareSymbolFont{symbols}{OMS}{cmss}{m}{n}
\DeclareSymbolFont{largesymbols}{OMX}{cmss}{m}{n}

\usepackage{graphicx}
\usepackage{subfigure}
\usepackage[all]{xy}

\begin{document}

\title{Towards First-principles Electrochemistry}

\author{Ismaila Dabo}
\email{daboi@cermics.enpc.fr}
\affiliation{CERMICS, Projet Micmac ENPC-INRIA, Universit\'e Paris-Est,
Marne-la-Vall\'ee, France}
\author{\'Eric Canc\`es}
\affiliation{CERMICS, Projet Micmac ENPC-INRIA, Universit\'e Paris-Est,
Marne-la-Vall\'ee, France}
\author{Yanli Li}
\affiliation{CERMICS, Projet Micmac ENPC-INRIA, Universit\'e Paris-Est,
Marne-la-Vall\'ee, France}
\author{Nicola Marzari}
\affiliation{Department of Materials Science and Engineering,
Massachusetts Institute of Technology, Cambridge, USA }
\affiliation{Department of Materials, University of Oxford, Oxford, United Kingdom}

\begin{abstract}
Chemisorbed molecules at a fuel cell electrode are a very sensitive probe of the
surrounding electrochemical environment, and one that can be accurately
monitored with different spectroscopic techniques. 
We develop a comprehensive electrochemical model to study molecular
chemisorption at either constant charge or fixed applied voltage, and calculate from
first principles the voltage dependence of vibrational frequencies---the vibrational Stark effect---for CO adsorbed
on close-packed platinum electrodes. The predicted vibrational Stark slopes are found to be in
very good agreement with experimental electrochemical spectroscopy data, thereby resolving previous controversies in
the quantitative interpretation of {\it in-situ} experiments
and elucidating the relation between canonical and grand-canonical descriptions of vibrational surface phenomena.
\end{abstract}

\maketitle

\section{Introduction}

Rising sustainability concerns have revived strong interest in electrochemical
electricity generation \cite{TesterDrake2005} whose basic principle is to 
catalytically convert the energy stored in chemical
bonds into usable electrical power. For any given electrochemical system, the 
power generated is the product of
two distinct contributions: the electrode voltage difference, which is the thermodynamic variable that 
quantifies the energy per electron made available through the breaking and rearranging of chemical bonds---Nernst's law---,
and the current density, which is the kinetic observable that 
measures the rate at which these chemical processes take place---Arrhenius' law. 
It should be noted, however, that
these two factors are not completely independent,
as one observes experimentally a systematic drop in voltage
at high electrical current. 
This  phenomenon, commonly known as activation voltage loss, 
represents one of the main limitations to the performance of electrochemical 
technologies \cite{LarminieDicks2003}.

Although the origins of the voltage dependence of the electrical current have long been conceptually understood \cite{Schmickler1996},
it is only recently that computational laboratories have applied first-principles
calculations to study this effect with the difficult task of describing catalytic
reactions at an electrode surface as a function of the applied voltage. 
The electrochemical free-energy correction introduced by N{\o}rskov {\it et al.} 
\cite{NorskovRossmeisl2004,KarlbergJaramillo2007}
represents a key successful step in this direction. In this
approach, the influence of the electrode voltage ${\cal E}$ is  included by adding
a correction $-e{\cal E}$ to the energy 
of all reaction intermediates that involve an electron
transferred to the metal. This zeroth-order correction has been shown to
accurately predict the activation voltage
of fundamental electrocatalytic processes, such 
as the oxygen reduction reaction at fuel-cell cathodes \cite{NorskovRossmeisl2004}.
However, this approach does not capture the self-consistent
modifications of the electronic structure
that arise from the applied potential. In particular, it does not consider
the variation of the electrode charge as a function of the potential and the interaction of 
the induced surface electric field with chemisorbed molecules.

Several other authors have proposed to account for these important electronic effects using more representative
electrochemical models \cite{LozovoiAlavi2001,LozovoiAlavi2003,OtaniSugino2006,TaylorWasileski2006,KarlbergRossmeisl2007,JinnouchiAnderson2008}.
These calculations differ in key quantitative and qualitative details 
from the approach we present, and are generally carried out at constant electrode 
charge $q$, the voltage $\cal E$ being determined at the post-processing stage using 
various procedures to relate  $\cal E$ to the computed Fermi level $\epsilon_{\rm F}$.
This is in contrast to typical experiments, in which the state of the system is directly controlled
via the electrode voltage and all relevant electrochemical properties are given in
terms of this central intensive variable. Although the voltage dependencies of
electrochemical properties can be recovered via an inverse Legendre transform, 
this indirect method entails repeated constant-charge calculations
to invert the charge-to-voltage relation. In addition, while canonical and
grand-canonical surface descriptions become equivalent in the limit of macroscopic systems, they can
significantly differ in describing microscopic phenomena. Indeed, although one can safely employ a
constant-charge framework for predicting surface phenomena that globally modify the electrical state of
the whole electrode (e.g., voltage-induced collective desorption, surface reconstruction), one should
systematically verify the validity of the constant-charge approximation when studying local electrochemical phenomena 
that do not affect the state of the entire electrochemical system (e.g., local surface reaction).

In this study, we introduce a practical computational model 
that allows, in particular, to work directly at fixed electrode voltage while fully describing self-consistent
changes in the electronic structure and taking into account realistic electrochemical conditions. 
A validation of the method
is provided by the prediction of the 
voltage dependence of vibrational frequencies---the vibrational Stark effect---for
 CO on platinum electrodes, which offers a very sensitive
spectroscopic probe of the electrochemical
environment and surface electric field. 
Since the vibrational properties of chemisorbed molecules can be accurately described from first principles
\cite{DaboWieckowski2007} and precisely measured using various infrared techniques \cite{MaillardLu2005,LuLagutchev2005}, the calculation of
the Stark effect represents a stringent test for assessing
the predictive ability of an electrochemical model. 
To date, notable discrepancies 
between first-principles predictions and
Stark measurements have been reported,
and their elucidation has been 
an important question in surface science and electrochemistry \cite{LozovoiAlavi2007,DeshlahraWolf2009}.

\section{Chemistry under applied electrode voltage}

\begin{figure}[ht]
\includegraphics{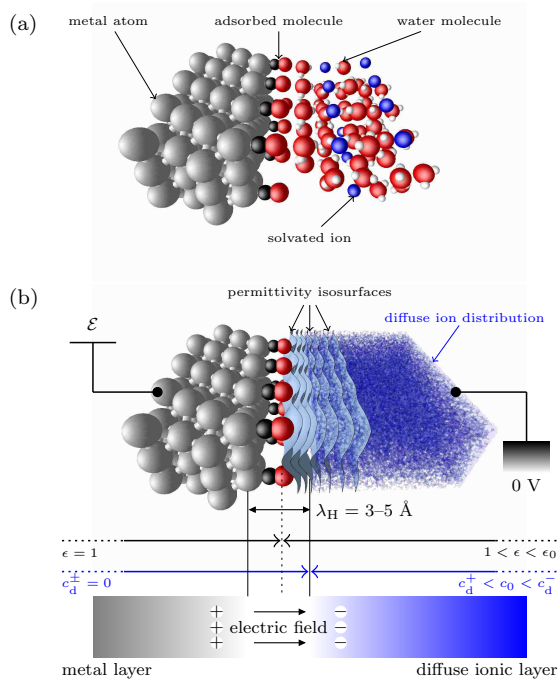}
\caption{(a) Adsorbate-covered catalytic electrode-electrolyte interface and (b) implicit
atom-continuum model of the double layer interface at fixed electrode voltage.}
\label{ElectrochemicalModel}
\end{figure}

To put matters into perspective, a typical electrode-electrolyte interface is depicted
in Fig. \ref{ElectrochemicalModel}(a).
The system consists of an adsorbate-covered metal surface in contact with an electrolyte solution.
For this system, the electrode voltage ${\cal E}$  corresponds to the energy involved in displacing an electron
from the metal electrode to the 
bulk of the ionic solvent \cite{BockrisKhan1993}. 
Therefore, including voltage conditions requires
to accurately describe the behavior of the screened electrostatic potential in the solvent region. 
It is important to note, however, that electrostatic screening in the electrolyte occurs on
considerably large length scales---typically, 10--$10^3$ \AA\
for ionic concentrations in the range  $10^{-4}$--$10^{-1}$ mol/liter (M) \cite{Sato1998}---, \linebreak
which renders the
explicit first-principles representation of the electrolyte
prohibitively expensive and practically inaccessible with current computational resources in the dilute electrolyte limit. 
Furthermore, quantitative errors in local and semilocal density-functional theory (DFT) descriptions
of water \cite{GrossmanSchwegler2004} have been reported. These errors result in
overstructured representations of aqueous media and overestimated freezing temperatures \cite{SitMarzari2005}---which
translates into overevaluated dielectric responses for explicit solvation models. 

To overcome these important limitations, we introduce an atom-continuum double layer model of the 
electrified interface [Fig. \ref{ElectrochemicalModel}(b)] \cite{Dabo2007,DaboBonnet2010}.
This model consists of immersing the metal electrode (the explicit metal layer)
in a semi-infinite electrolytic continuum (the implicit diffuse layer) \cite{BockrisKhan1993}. 
Note that the thickness $\lambda_{\rm H}$ of the interphase region (the Helmholtz interface)
is experimentally found to be equal to 3--5 \AA\ (that is, approximately the thickness of a water bilayer) 
regardless of the nature of the electrode compound \cite{Sato1998}.
Within this implicit approach, the electrode voltage ${\cal E}$ can be simply expressed as
\begin{equation}
{\cal E} = {\cal V}_{0}- \frac{1}{e}\epsilon_{\rm F} = - \frac{1}{e}\epsilon_{\rm F},
\label{VoltageDefinition}
\end{equation}
where  ${\cal V}_{\rm 0}=0$ V is the asymptotic value of the potential in the bulk of the electrolyte.
The above expression of the electrode voltage ${\cal E}$ is obtained from
the conventional definition of the absolute electrode potential 
in terms of single-electron electrical work and chemical energy differences \cite{BockrisReddy1998,DaboBonnet2010}.
Alternative definitions of the electrode voltage have been proposed; 
we refer the reader to Refs. \onlinecite{Trasatti1990,Pethica2007,Fawcett2008} for critical discussions
of electrode potential concepts. 

With this physical picture in mind, it clearly appears that increasing the electrode voltage
causes a depletion of surface electronic states
compensated by an accumulation of negative counterions in the electrolyte [Fig. \ref{ElectrochemicalModel}(b)].
The induced polarization enhances the double layer electric field,
which interacts more strongly with the adsorbates
and shifts their vibrational frequencies.

In this model, a smoothly varying
dielectric permittivity $\epsilon$ accounts for the water environment, and 
diffuse charge densities $c_{\rm d}^+$ and $c_{\rm d}^-$ represent the
thermal distributions of the counterions of bulk concentration $c_0$, 
absolute charge $z_{\rm d}$, and size $a_{\rm d}$
(for the sake of simplicity, we restrict ourselves
to the case of a $z_{\rm d}$:$z_{\rm d}$ symmetric ionic solution).
The dielectric permittivity is calculated using the parameterization of 
Gygi and Fattebert \cite{FattebertGygi2002}, which involves
the static dielectric constant of water $\epsilon_0=78$ using a smeared superposition
of atomic electronic densities $\tilde \rho({\bf r})$ to define the local permittivity
\begin{equation}
\epsilon({\bf r}) = 1 + \frac{\epsilon_0-1}{2} \left(
1 + \frac{1-\left({\tilde \rho({\bf r})}/{\rho_0}\right)^{2\beta}}
{1+\left({\tilde \rho({\bf r})}/{\rho_0}\right)^{2\beta}}
\right),
\label{LocalPermittivity}
\end{equation}
In Eq. (\ref{LocalPermittivity}), the threshold density $\rho_0$ sets the solvation shell 
and the exponent $\beta$ defines the smoothness of the dielectric
transition. It is important to note that we use here a fixed superposition of atomic densities $\tilde \rho({\bf r})$ in place of the 
self-consistent charge density $\rho({\bf r})$ to cancel spurious surface contributions to the one-electron potential 
and to prevent the electronic charge from flowing into the electrolyte \cite{SanchezSued2009,DaboBonnet2010}. 
A possible alternative to this approach is to use a real-space preconditioning of the total energy gradient \cite{Dabo2007,DaboBonnet2010}.
This dielectric model  properly captures the gradual transition of the
permittivity across the solvation shell \cite{MarenichCramer2008} and yields accurate solvation
energies for a broad range of molecular species with proper parameterization \cite{ScherlisFattebert2006,SanchezSued2009}. 
The ionic concentrations $c_{\rm d}^+$ and 
$c_{\rm d}^-$ follow the modified Boltzmann statistics introduced by Borukhov, Andelman, and Orland
\cite{BorukhovAndelman1997}, which includes finite-size steric interactions between counterions.
Note that the contribution from explicit water overlayers
can always be included. However,
due to weak chemical interactions between CO and the solvent \cite{RoudgarGross2005},
 we restrict here the explicit DFT treatment to the metal and chemisorbed molecules.

The ground state of the electrochemical system is that which minimizes
the free energy functional
\begin{equation}
G = E' + \Delta E^{\rm corr} + \Delta E^{\rm ion} - {\cal E}q.
\label{TotalEnergy}
\end{equation}
The first contribution $E'$ corresponds to the DFT energy of the system
within the supercell approximation (that is, for a periodically repeated metal slab
in vacuum), as computed by standard plane-wave codes \cite{PayneTeter1992}.
The corrective energy $\Delta E^{\rm corr}$  equals the difference 
between the electrostatic energy of the isolated slab
and that of the periodic slab in vacuum,
\begin{equation}
\Delta E^{\rm corr} = \frac{1}{2} \int \rho({\rm r})  v^{\rm corr}({\rm r}) d{\bf r},
\label{CorrectiveEnergy}
\end{equation}
where $\rho$ is the explicit charge density, 
and $v^{\rm corr}=v-v'$ is a corrective potential defined as 
the difference between 
the Coulomb potential of the solvated system $v$ that
satisfies a nonlinear modified Poisson-Boltzmann (MPB) equation
\begin{equation}
\nabla \cdot \epsilon({\bf r}) \nabla v({\bf r}) = - 4 \pi [ \rho({\bf r}) - z_{\rm d} c^+_{\rm d}({\bf r}) + z_{\rm d} c^-_{\rm d}({\bf r}) ]
\label{MPBE}
\end{equation}
(in a.u.) with boundary conditions $v=0$ V at infinity,
and the periodic potential $v'$ calculated in the reciprocal space representation
using fast Fourier transform techniques \cite{PayneTeter1992}.
Heuristically, $v^{\rm corr}$ can be identified as the electrolyte reaction field
\cite{MarenichCramer2008}, which self-consistently
accounts for the influence of the applied electrode voltage.
The remaining contribution $\Delta E^{\rm ion}$
is that from the counterions in solution, including steric repulsion \cite{BorukhovAndelman1997}:
\begin{eqnarray}
\Delta E^{\rm ion} =  \displaystyle \int  d {\bf r} \left\{ \displaystyle\frac{1}{2}
z_{\rm d} ( c_{\rm d}^- - c_{\rm d}^+ ) v - ( c_{\rm d}^+ +  c_{\rm d}^- - 2 c_0) 
\mu - T \left[ s^{\rm ion}(c_{\rm d}^+,c_{\rm d}^-,c_{\rm d}^\circ) - s^{\rm ion}(c_0,c_0,a_{\rm d}^{-3})\right] \right\},
\end{eqnarray}
where $\mu= - k_{\rm B} T \ln( a_{\rm d}^{-3} c_0^{-1} - 2)$
is the ionic potential, $s^{\rm ion}$ is the local ionic entropy, 
and $c_{\rm d}^\circ$ is the saturated (maximum packing) ionic concentration 
that smoothly goes from 0 to $a_{\rm d}^{-3}$ 
at a distance $\lambda_{\rm H}$ from the metal surface.
The entropy density $s^{\rm ion}$ can be expressed as
\begin{eqnarray}
s^{\rm ion}(c_{\rm d}^+,c_{\rm d}^-,c_{\rm d}^\circ) 
= - k_{\rm B}\left[ c_{\rm d}^+ \ln \left( \frac{c_{\rm d}^+}{c_{\rm d}^\circ} \right)
+ c_{\rm d}^- \ln \left( \displaystyle\frac{c_{\rm d}^-}{c_{\rm d}^\circ} \right) 
+ ( c_{\rm d}^\circ -  c_{\rm d}^+ -  c_{\rm d}^- )
\ln \left( 1 - \displaystyle\frac{c_{\rm d}^+}{c_{\rm d}^\circ} 
- \displaystyle\frac{c_{\rm d}^-}{c_{\rm d}^\circ} \right)  \right] ,
\end{eqnarray}
and the maximal concentration $c_{\rm d}^\circ({\bf r})$ is parameterized as
\begin{equation}
c_{\rm d}^\circ({\bf r}) = \frac{1}{a_{\rm d}^3} 
\left. \prod_{I} \right. ^\prime \widetilde \Theta
\left( |{\bf r} - {\bf R}_I| - \lambda_{\rm H}  \right),
\end{equation}
where the index $I$ runs exclusively over the metal layer atoms located at positions ${\bf R}_I$, 
and the counterion exclusion region is defined by a smooth step
function $\widetilde \Theta$ smeared over a limited number of grid points for numerical convergence.
Consequently, 
the equilibrium ionic concentrations read
\begin{eqnarray}
 c_{\rm d}^{\pm}({\bf r}) & =  & c_{\rm d}^\circ({\bf r}) \exp\left(\pm \displaystyle\frac{ z_{\rm d} v({\bf r})}{k_{\rm B}T}\right) 
\left\{ 
c_0^{-1} a_{\rm d}^{-3} + 2 \left[ \cosh\left( 
\displaystyle\frac{ z_{\rm d} v({\bf r})}{k_{\rm B}T}
\right) - 1 \right] \right\}^{-1}.
\label{IonicConcentration}
\end{eqnarray}
(It is important to note that the ionic distribution defined by Eq. (\ref{IonicConcentration})
approach Gouy-Chapman-Stern distributions in the limit of small ionic sizes.)
This ionic model directly involves the Helmholtz thickness $\lambda_{\rm H}$ through 
the prefactor $c_{\rm d}^\circ$ and
provides a simple representation of the diffuse distributions in direct connection to
the Stern picture \cite{BockrisKhan1993}.

\section{Implementation of the electrochemical model and results}

The implementation of this electrochemical model raises three main difficulties.
First, solving the MPB electrostatic problem in a finite simulation cell with,
e.g., periodic or homogeneous boundary conditions results in significant errors in the electrode
voltage that increase exponentially with ionic dilution. To correctly extrapolate the slowly vanishing electrostatic potential, we impose
fictitious electrochemical boundary conditions
obtained from the long-range integration of the MPB equation in the 
planar-average approximation:
\begin{equation}
\nabla v \cdot {\bf n}_z=-\sqrt{\frac{32 \pi c_0 k_{\rm B} T}{\epsilon_0}}
\sinh \frac{z_{\rm d} v}{2k_{\rm B}T},
\end{equation}
where ${\bf n}_z$ denotes the external surface vector.

Additionally, the solution of the MPB equation is 
expensive due to the fine grids required in discretizing the charge density. 
To reduce this computational burden, we exploit the fact that the corrective potential
$v^{\rm corr}$ varies smoothly over space, which allows for its inexpensive calculation in the spirit of the density-countercharge method 
\cite{DaboKozinsky2008} by solving the simplified corrective potential equation
\begin{equation}
\nabla \cdot \epsilon({\bf r}) \nabla v^{\rm corr}({\bf r}) = - 4 \pi [ \langle \rho \rangle + 
\rho'_{\rm p}({\bf r}) - z_{\rm d} c^+_{\rm d}({\bf r}) + z_{\rm d} c^-_{\rm d}({\bf r}) ]
\label{CorrectiveMPBE}
\end{equation}
where
\begin{equation}
\rho'_{\rm p}({\bf r})= \frac{1}{4\pi} \nabla \cdot (\epsilon({\bf r})-1) \nabla v'({\bf r})
\end{equation}
is the interfacial polarization density induced by the vacuum periodic solution of 
the electrostatic problem $v'({\bf r})$ that satisfies
\begin{equation}
\nabla^2 v'({\bf r}) = - 4 \pi ( \rho({\bf r}) - \langle \rho \rangle ).
\end{equation}
Note, in particular, that the source term of
Eq. (\ref{CorrectiveMPBE}) does not explicitly involve the charge density $\rho$ and
is much smoother than that entering into the original electrostatic problem [Eq. (\ref{MPBE})],
allowing for the computation of the corrective solvation potential $v^{\rm corr}$ on coarse-grained meshes \cite{Dabo2007,DaboBonnet2010}
using, e.g.,  multigrid solvers \cite{HolstSaied1993,HolstSaied1995}.

Last, constant-potential simulations require fixing the Fermi energy while readjusting
the electron number during the electronic-structure optimization. 
In the course of such calculations,
large charge oscillations occur, which results in systematic energy divergence. 
To eliminate these instabilities without 
resorting to artificial charge compensations, we employ a generalization of
the ensemble density-functional theory scheme \cite{MarzariVanderbilt1997} 
and optimal damping algorithm  \cite{CancesLeBris2000}, which ensures that the free energy converges monotonically. 
In our current finite-temperature implementation, the line minimization of the one-particle density matrix represents a substantial 
fraction of the computational effort. The optimization of this finite-temperature line-search procedure 
will be important for future applications. 

\begin{figure}[ht]
\includegraphics{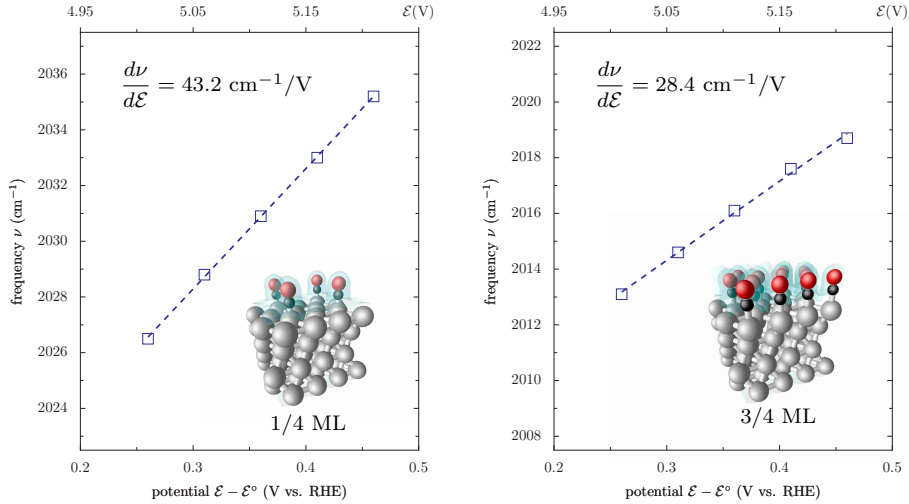}
\caption{Vibrational frequency $\nu$ as a function of the NHE-referenced electrode voltage ${\cal E}-{\cal E}^\circ$ for
1/4 ML and 3/4 ML of CO on Pt(111). The absolute electrode potential $\cal E$ is also indicated.}
\label{StarkTuning}
\end{figure}

We thus proceed to calculate the vibrational properties of CO-covered
platinum interfaces under electrochemical 
conditions. In carrying out these calculations,
we employ the Perdew-Burke-Ernzerhof semilocal density-functional
approximation \cite{PerdewBurke1996} and ultrasoft pseudopotentials \cite{Vanderbilt1990}. The Brillouin zone is sampled with
a shifted 4 $\times$ 4 $\times$ 1 reciprocal-space integration grid. Plane-wave kinetic energy cutoffs of 25 and 200 Ry are applied to
the Fourier expansion of the wavefunctions and electronic charge density, respectively.
The system is represented by fully relaxed three-layer-thick Pt(111) slabs
at the calculated bulk lattice parameter of 3.99 \AA\
in a $\sqrt{3} \times 2$ supercell. 
The solvation parameters are set to the values determined and used in Ref. \onlinecite{ScherlisFattebert2006}, i.e.,
$\rho_0=0.00078$ a.u. and $\beta=1.3$. We impose to the density dependence of the dielectric permittivity to drop quadratically
 to $\epsilon=1$ below the threshold value $\epsilon=1.5$ to eliminate numerical fluctuations in $\epsilon$
in the electrode region. A Gaussian spread of 1 bohr is used to smear the atomic superposition
$\tilde \rho$. The counterion exclusion region is defined using a step
function $\widetilde \Theta$ smeared over $\sim$0.5 bohr.
The electrode voltage range, counterion concentration, and ionic temperature are selected to be ${\cal E}=5.0$--5.2 V,
$c_0=0.1$ M, and $T=300$ K, respectively, corresponding to experimental conditions. 
We reference the absolute potential of the half cell
to that of the normal hydrogen electrode (NHE) by matching the electrode voltage calculated
at the point of zero charge ${\cal E}_{\rm pzc}=5.07$ V
to the referenced experimental potential ${\cal E}_{\rm pzc}^{\rm exp}-{\cal E}^\circ
=0.33$ V \cite{GomezCliment2000}. This corresponds to shifting
the absolute electrode potential by ${\cal E}^\circ=4.74$ V in our calculations.
The thickness of the 
double layer $\lambda_{\rm H}$ equals 4 \AA, that is, in the middle of the experimental range 3--5 \AA.
The size and charge of the counterions 
are chosen to be $a_{\rm d}=2$ \AA\ and $z_{\rm d}=1$ for a typical monovalent electrolyte. 
We compute stretching frequencies using a frozen-phonon method with vertical 
atomic displacements of a few hundreds of bohr and a bicubic fit of the two-dimensional
potential energy surface. The calculated
frozen-phonon frequencies agree to within 1--2 cm$^{-1}$ to the results
of the full computation and diagonalization of the dynamical matrix \cite{DaboWieckowski2007}.

The dependency of the C--O intramolecular frequencies 
as a function of the electrode voltage at surface concentrations of 1/4 and 3/4 ML for the experimentally
observed atop configuration is depicted in Fig. \ref{StarkTuning}. (Note that local and semilocal calculations
fail to predict the relative stability of CO adsorption sites on close-packed platinum electrodes
\cite{FeibelmanHammer2001,DaboWieckowski2007} due to self-interaction overhybridization errors \cite{KresseGil2003,DaboFerretti2010}.)
First, we observe that the predicted vibrational frequencies follow a common
increasing and almost linear trend as a function of the electrode potential,
in qualitative agreement with experiment under
preoxidation conditions, i.e., below $\sim$0.5 V vs. NHE. Nevertheless, 
it should be observed that the C--O stretching frequency drops sensibly with increasing
CO monolayer coverage at variance with experimental observations \cite{DeshlahraWolf2009}. The origin of this frequency shift 
is still not entirely elucidated (at this stage, 
we ascribe this effect to unphysical lateral interaction with the implicit solvent at low surface coverage).

The vibrational Stark effect is seen to strongly depend on 
the CO monolayer concentration. Indeed, at a coverage of 1/4 ML, the vibrational Stark slope is calculated to be 
$d\nu/d{\cal E}= 43.2$ cm$^{-1}$/V, whereas at 3/4 ML, 
$d\nu/d{\cal E}$ equals 28.4 cm$^{-1}$/V. 
Despite this marked dependency on monolayer coverage, 
the Stark rates are in remarkable accordance with their experimental counterparts
$d\nu/d{\cal E}= 40$ cm$^{-1}$/V at 1/4 ML \cite{WeaverZou1999} and 
28 cm$^{-1}$/V at 3/4 ML \cite{MaillardLu2005,LuLagutchev2005}. 
Thus, the calculated Stark tuning rates are found here to 
deviate by less than 8\% from experimental spectroscopic measurements in both
the low- and high-coverage regimes, providing an important illustration
of the predictive performance of the present electrochemical model.
By performing a sensitivity analysis, we find the 
Stark slopes to be altered by less than 3 cm$^{-1}$/V when varying the main experimental parameter $\lambda_{\rm H}$ by $\pm$1 \AA.

Having validated our electrochemical model, we finally discuss the equivalence between canonical and grand-canonical
vibrational surface models. To this end, we have compared the results of
C--O stretching frequency calculations in the constant-voltage and constant-potential regimes for both positive and negative surface charges,
finding agreement of 1--3 cm$^{-1}$/V in the predicted Stark shifts. 
These results lend quantitative support to the view that 
adsorbate stretching modes represent a relatively weak perturbation of the state of the electrode surface and confirm
the equivalence of constant-charge and constant-voltage descriptions at low vibrational amplitude. 

\section{Summary and outlook}

In summary, we have developed a practical and comprehensive electrochemical model to study 
quantum-mechanical systems at both constant charge and fixed applied voltage.
We have used this model to describe the vibrational Stark effect for CO adsorbed on platinum
 surfaces, which represents an important probe of electrical conditions at the electrode-electrolyte interface. 
The calculated Stark tuning coefficients are found to be in very good
agreement with spectroscopic experiments at both low and high surface coverages. 

These results confirm the possibility of describing electrochemical conditions
with a simplified continuum model of the solvent and establish the 
predominance of electrostatic interfacial capacitance effects \cite{Schmickler1996,Dabo2007}
in predicting vibrational Stark shifts. In addition, the direct comparison between constant-voltage and constant-charge calculations
confirms the thermodynamical equivalence between the grand-canonical and canonical approaches for
low-amplitude vibrational surface phenomena. The
verification of this thermodynamical equivalence for the study of electrochemical surface reactions remains a question of central interest. 

Previous studies have shown that models relying on imposing an external surface electric field in a double
layer of typical thickness 3--5 \AA\ are not sufficiently refined for the reliable representation of the nonuniform surface electric
field \cite{LozovoiAlavi2007,DeshlahraWolf2009}. 
Our study demonstrates that close agreement with experiment can be achieved using a 
comprehensive implicit representation of the Helmholtz layer and ionic solvent.
These results resolve inconsistencies in the interpretation of electrochemical spectroscopic measurements
and open promising perspectives for the first-principles description of
electrocatalytic processes under electrochemical conditions.

\section{Acknowledgments}

The calculations in this work have been performed using the 
Quantum-Espresso package \cite{QuantumEspresso2009}. 
The authors acknowledge support from the MURI Grant DAAD 19-03-1-0169 and Sire Grant 06-CIS6-014 of the French National Agency of Research. 
We thank Andrzej Wieckowski for his help in providing and interpreting spectroscopic data. 
Helpful suggestions and comments from Gerbrand Ceder, Stefano de Gironcoli, Damian Scherlis, 
Nic\'ephore Bonnet, Jean-S\'ebastien Filhol, Michiel Sprik, and Eduardo Lamas are gratefully acknowledged.

\bibliography{article}

\end{document}